\input mtexsis


\newif\ifLocalProcessing\newif\ifForceTenPointFonts	 
\def\ATunlock{\catcode`@=11} \def\ATlock{\catcode`@=12} \ATunlock
\def\l@m{>\space} \def\t@m{....} \def\sp@ce{\space\space\space}
\immediate\write16{} \immediate\write16{ \l@m Testing for the TeX dialect\t@m}

%
\ifx\@latexerr\und@fin@d\relax				
\else							
  \newlinechar=`\^^J
  \immediate\write16{
    \sp@ce What?? LaTeX!! LaTeX?!?!?!^^J
    ^^J
    \l@m You cannot print this document with LaTeX.^^J
    \l@m ^^J
    \l@m This preprint uses the TeXsis macro package. If TeXsis is installed^^J
    \l@m on your system, please use the command "texsis", not "latex".^^J
    \l@m ^^J
    \l@m You can also print it with Plain TeX. For that, you need the macro^^J
    \l@m file "mtexsis.tex". If you have this file, please run the command^^J
    \l@m "tex", not "latex"; if you do not have it, please get it from^^J
    \l@m hep-th, hep-lat, etc.; it is loaded automatically if you use Plain,^^J
    \l@m you need NOT edit the file \jobname.tex to add
         "\noexpand\input mtexsis".^^J
    \l@m ^^J
    \l@m If your printer is Postscript, I can also send you a Postscript^^J
    \l@m file of this preprint.^^J
    \l@m ^^J
    \l@m --Anil Trivedi (trivedi@yukawa.uchicago.edu)
    ^^J
                    }
\@@end\fi

%
\newread\@mtxfile
\ifx\texsis\und@fin@d					
  \newlinechar=`\^^J
  \immediate\write16{
    \sp@ce Not TeXsis.^^J^^J
    \l@m Looking for "mtexsis.tex"\t@m
                    }
  \immediate\openin\@mtxfile=mtexsis\relax		
  \ifeof\@mtxfile\closein\@mtxfile			
   \immediate\write16{\sp@ce Not found!!!!}
   \immediate\write16{
    ^^J
    \l@m This preprint uses the TeXsis macro package. To print it with^^J
    \l@m Plain TeX, you NEED the file mtexsis.tex (available from hep-th,^^J
    \l@m hep-lat, etc.). It is loaded automatically if you use Plain TeX,^^J
    \l@m you need NOT edit the file \jobname.tex to add
         "\noexpand\input mtexsis".^^J
    \l@m ^^J
    \l@m Of course, IF TEXSIS IS INSTALLED on your system, there is no^^J
    \l@m need for mtexsis.tex: just use the command "texsis", not "tex".^^J
    \l@m OTHERWISE, please get mtexsis.tex and run "tex \jobname" again.^^J
    \l@m ^^J
    \l@m If your printer is Postscript, I can also send you a Postscript^^J
    \l@m file of this preprint.^^J
    \l@m ^^J
    \l@m --Anil Trivedi (trivedi@yukawa.uchicago.edu)
    ^^J
                     }
    \end						
  \else\immediate\closein\@mtxfile			
 %
 %
  \immediate\write16{\sp@ce Loading in batchmode; please be patient.}
  \batchmode\input mtexsis\errorstopmode		
  \ATunlock\fi						
\else\immediate\write16{\sp@ce TeXsis.}\relax		
\fi



\def\f@n{\fontname}\def\g@f{\global\font}
  \def\m@{   scaled \magstep}
\def\m@hf{   scaled \magstephalf}

    \newif\ifoldfonts
    \edef\tenrmname{\fontname\tenrm}
    {\def\\#1{\catcode`#1=12 }\\a\\m\\r\\1\\0\gdef\AMRTEN{amr10}}  
    \ifx\tenrmname\AMRTEN\oldfontstrue\fi

%
\def\everyF@NT{\relax}				
\def\F@NT#1=#2#3#4#5#6{
    \ifLocalProcessing				
    \g@f#1=#2
    \else%
       \ifoldfonts
       \batchmode\g@f#1=#5\errorstopmode
       \else%
          \ifForceTenPointFonts
          \batchmode\g@f#1=#4\errorstopmode
          \else%
             \batchmode\g@f#1=#2\errorstopmode
             \ifx#1\nullfont%
             \batchmode\g@f#1=#3\errorstopmode
             \ifx#1\nullfont%
             \batchmode\g@f#1=#4\errorstopmode
             \fi\fi%
          \fi
          \ifx#1\nullfont%
             \batchmode\g@f#1=#5\errorstopmode
          \fi%
       \fi%
       \ifx#1\nullfont%
          \message{\sp@ce #1\def'ed to #6(desperate choice!)}
          \gdef#1{#6}
       \else%
          \message{\sp@ce #1set as \f@n#1}%
       \fi%
       \everyF@NT				
    \fi%
                      }

\emsg{}\emsg{ \l@m Fonts\t@m}%
\ifLocalProcessing\emsg{\sp@ce(9, 10, 12, 14, 16 pt.)}
\else%
  \ifoldfonts
  \emsg{\sp@ce(your \NX\tenrm was amr10; I'll only look for old fonts)}\emsg{}%
  \else%
     \ifForceTenPointFonts%
     \emsg{\sp@ce(you asked for magnified 10pt fonts)}\emsg{}%
     \else%
     \emsg{\sp@ce(find the closest size, then scale if necessary)}%
     \fi%
  \fi
\fi%

\def\ninefonts{
  \F@NT\ninerm={cmr9}{}{cmr9}{amr9}{\tenrm}%
  \F@NT\ninei={cmmi9}{}{cmmi9}{ammi9}{\teni}%
  \F@NT\ninesy={cmsy9}{}{cmsy9}{amsy9}{\tensy}%
  \F@NT\nineex={cmex10}{}{cmex10}{amex10}{\tenex}
  \F@NT\ninebf={cmbx9}{}{cmbx9}{ambx9}{\tenbf}%
  \F@NT\ninesl={cmsl9}{}{cmsl9}{amsl9}{\tensl}%
  \F@NT\ninett={cmtt9}{}{cmtt9}{amtt9}{\tentt}%
  \F@NT\nineit={cmti9}{}{cmti9}{amti9}{\tenit}%
  \skewchar\ninei='177\skewchar\ninesy='60\hyphenchar\ninett=-1%
  \gdef\ninefonts{\relax}%
              }

\F@NT\tenss={cmss10}{}{cmss10}{amss10}{\tenrm}
\F@NT\tencsc={cmcsc10}{}{cmcsc10}{amcsc10}{\teni}
\let\tenp@int=\tenpoint
\def\tenpoint{\tenp@int\def\sc{\tencsc}\def\csc{\sc}}

\def\twelvefonts{
  \F@NT\twelverm={cmr12}{cmr10\m@1}{cmr10\m@1}{\f@n\tenrm\m@1}{\twelvess}
  \F@NT\twelvei={cmmi12}{cmmi10\m@1}{cmmi10\m@1}{\f@n\teni\m@1}{}
  \F@NT\twelvesy={cmsy10\m@1}{cmsy10\m@1}{cmsy10\m@1}{\f@n\tensy\m@1}{}
  \F@NT\twelveex={cmex10\m@1}{cmex10\m@1}{cmex10\m@1}{\f@n\tenex\m@1}{}
  \F@NT\twelvebf={cmbx12}{cmbx10\m@1}{cmbx10\m@1}{\f@n\tenbf\m@1}{\twelverm}
  \F@NT\twelveit={cmti12}{cmti10\m@1}{cmti10\m@1}{\f@n\tenit\m@1}{\twelvesl}
  \F@NT\twelvesl={cmsl12}{cmsl10\m@1}{cmsl10\m@1}{\f@n\tensl\m@1}{\twelveit}
  \F@NT\twelvett={cmtt12}{cmtt10\m@1}{cmtt10\m@1}{\f@n\tentt\m@1}{\twelverm}
  \F@NT\twelvess={cmss12}{cmss10\m@1}{cmss10\m@1}{amss10\m@1}{\twelverm}
  \F@NT\twelvecsc={cmcsc10\m@1}{}{cmcsc10\m@1}{amcsc10\m@1}{\twelveit}%
  \skewchar\twelvei='177\skewchar\twelvesy='60\hyphenchar\twelvett=-1%
  \gdef\twelvefonts{\relax}%
                }
\let\twelvep@int=\twelvepoint
\def\twelvepoint{\twelvep@int\def\sc{\twelvecsc}\def\csc{\sc}}

\def\fourteenfonts{
  \F@NT{\fourteenrm}={cmr12\m@1}{cmr10\m@2}{cmr10\m@2}{\f@n\tenrm\m@2}{}
  \F@NT{\fourteeni}={cmmi12\m@1}{cmmi10\m@2}{cmmi10\m@2}{\f@n\teni\m@2}{}
  \F@NT{\fourteensy}={cmsy10\m@2}{cmsy10\m@2}{cmsy10\m@2}{\f@n\tensy\m@2}{}
  \F@NT{\fourteenex}={cmex10\m@2}{cmex10\m@2}{cmex10\m@2}{\f@n\tenex\m@2}{}
  \F@NT{\fourteenbf}={cmbx12\m@1}{cmbx10\m@2}{cmbx10\m@2}{\f@n\tenbf\m@2}{}
  \F@NT{\fourteenit}={cmti12\m@1}{cmti10\m@2}{cmti10\m@2}{\f@n\tenit\m@2}{}
  \F@NT{\fourteensl}={cmsl12\m@1}{cmsl10\m@2}{cmsl10\m@2}{\f@n\tensl\m@2}{}
  \F@NT{\fourteenss}={cmss12\m@1}{cmss10\m@2}{cmss10\m@2}{amss10\m@2}{}
  \skewchar\fourteeni='177\skewchar\fourteensy='60%
  \gdef\fourteenfonts{\relax}%
                  }

\def\sixteenfonts{
  \F@NT{\sixteenrm}={cmr17}{cmr12\m@2}{cmr10\m@3}{\f@n\tenrm\m@3}{}
  \F@NT{\sixteeni}={cmmi12\m@2}{cmmi10\m@3}{cmmi10\m@3}{\f@n\teni\m@3}{}
  \F@NT{\sixteensy}={cmsy10\m@3}{cmsy10\m@3}{cmsy10\m@3}{\f@n\tensy\m@3}{}
  \F@NT{\sixteenex}={cmex10\m@3}{cmex10\m@3}{cmex10\m@3}{\f@n\tenex\m@3}{}
  \F@NT{\sixteenbf}={cmbx12\m@2}{cmbx10\m@3}{cmbx10\m@3}{\f@n\tenbf\m@3}{}
  \F@NT{\sixteenit}={cmti12\m@2}{cmti10\m@3}{cmti10\m@3}{\f@n\tenit\m@3}{}
  \F@NT{\sixteensl}={cmsl12\m@2}{cmsl10\m@3}{cmsl10\m@3}{\f@n\tensl\m@3}{}
  \F@NT{\sixteenss}={cmss17}{cmss12\m@2}{cmss10\m@3}{amss10\m@3}{}
  \F@NT{\sixteenssi}={cmssi17}{cmssi12\m@2}{cmssi10\m@3}{amssi10\m@3}{}
  \skewchar\sixteeni='177\skewchar\sixteensy='60%
  \gdef\sixteenfonts{\relax}%
                 }
\sixteenpoint\twelvepoint   	

\emsg{}\emsg{ \l@m Looking for the script font "rsfs"\t@m}%
\newfam\Scrfam\newfam\scrfam				
\batchmode\font\scrfont=rsfs10\errorstopmode		
\ifx\scrfont\nullfont					
   \emsg{\sp@ce Not found!!!}  \emsg{}
   \emsg{ \l@m You don't seem to have the script font "rsfs"; get it!}
   \emsg{ \l@m For now, the substitution \noexpand\scr -> \noexpand\cal
            will be made.}
   \emsg{ \l@m --Anil Trivedi} \emsg{}
   \def\s@r{\cal}					
\else							
   \g@f\twelvescr=rsfs10\m@1 \g@f\eightscr=rsfs7\m@1 \g@f\sixscr=rsfs5\m@1
   \skewchar\twelvescr='177\skewchar\eightscr='177\skewchar\sixscr='177
   \textfont\Scrfam=\twelvescr\scriptfont\Scrfam=\eightscr
   \scriptscriptfont\Scrfam=\sixscr
   \g@f\tenscr=rsfs10 \g@f\sevenscr=rsfs7 \g@f\fivescr=rsfs5
   \skewchar\tenscr='177\skewchar\sevenscr='177\skewchar\fivescr='177
   \textfont\scrfam=\tenscr\scriptfont\scrfam=\sevenscr
   \scriptscriptfont\scrfam=\fivescr
   \def\scr{\fam\Scrfam}				
   \def\s@r{\scr}					
   \message{\sp@ce Script font loaded.}
\fi


\emsg{}\emsg{ \l@m Loading Macros\t@m}


\overfullrule=0pt\vbadness=10000\hbadness=10000\tolerance=400
\def\email#1{\smallskip\rm(#1)}

\def\setp@rskip#1{%
   \parskip=#1\baselineskip
   \parskip=\the\parskip plus 1pt minus 1pt}
\def\setparskip{\setp@rskip{.2}}			

\long\def\multiplyspacing#1{
   \baselineskip = #1\baselineskip
   \setRuledStrut\setTableskip				
   \setparskip}
\def\lsfactor#1{\singlespaced\multiplyspacing{#1}}
\def\normalspacing{\lsfactor{1.2}}			

\EnvTopskip=\smallskipamount\EnvBottomskip=\smallskipamount

\def\quotefont{\tenrm} \def\quotespacing{\normalspacing}


\def\It#1{{\it#1\/}}

\def\;{\relax\ifmmode\mskip\thickmuskip\else\,\,\fi}		

\def\jnl{\bgroup\catcode`\.=\active\offparens\@jnl}\offparens
\def\@jnl'#1'#2'#3'#4'{
   \ifEurostyle {#1} {\vol{#2}} (#4) #3\relax
   \else {#1} {\vol{#2}}, #3 (#4)\relax\fi\egroup
                      }
\newif\ifsuperrefs

\def\SetRefs{
   \def\Ref##1{Ref.~\l@ref\use{Ref.##1}\r@ref}	
   \def\ref##1{\l@ref\use{Ref.##1}\r@ref}	
   \def\xref##1{\use{Ref.##1}}			
   \def\citerange{\refrange}			
   \def\Refrange##1##2{
      Refs.~\l@ref\use{Ref.##1}--\use{Ref.##2}\r@ref}
   \ifsuperrefs					
      \def\l@ref{}\def\r@ref{}
   \else					
      \def\l@ref{[}\def\r@ref{]}
  \fi}
\let\sup@rr@fstrue=\superrefstrue
\let\sup@rr@fsfalse=\superrefsfalse
\def\superrefstrue{\sup@rr@fstrue\SetRefs}
\def\superrefsfalse{\sup@rr@fsfalse\SetRefs}
\superrefsfalse					

%
\def\Eq#1{Eq.~($\use{Eq.#1}$)}			
\def\Eqs#1{Eqs.~($\use{Eq.#1}$)}		
\def\Ep#1{($\use{Eq.#1}$)}          		
%
%
\def\@use#1{\endgroup\stripblanks @#1@\endlist
   \XA\ifx\csname\tok\endcsname\relax\relax
     \emsg{\l@m UNDEFINED TAG #1 ON PAGE \folio.}%
     \global\advance\@BadTags by 1
     \@errmark{UNDEF}\edef\tok{{\bf\tok}}%
   \else\edef\tok{\csname\tok\endcsname}%
   \fi\tok}%
\def\make@refmark#1{
   \testtag{Ref.#1}\ifundefined
     \emsg{\l@m UNDEFINED REFERENCE #1 ON PAGE \number\pageno.}%
     \global\advance\@BadRefs by 1
     \xdef\@refmark{{\tenbf #1}}\@errmark{REF?}
   \else\xdef\@refmark{\csname\tok\endcsname}%
   \fi}



\def\keepInformal{\def\Informal{\relax}\def\endInformal{\relax}}
\def\noInformal{\long\def\Informal##1\endInformal{\relax}
                \def\endInformal{\relax}}

\keepInformal							

\long\def\informalabstract#1\endinformalabstract{
   \Informal\informalabstractskip			
   \noindent(\It{Informal Abstract.}\ #1)
   \endInformal\def\endinformalabstract{\relax} }

\newif\iffirstpj\firstpjtrue
\long\def\pj#1\endpj{
   \Informal
   \iffirstpj\vskip0pt plus1filll\firstpjfalse
   \else\vskip1.5\baselineskip\fi
   \bgroup \def\\{\par}					
   \def\pjauthor{\twelvess\medskip---\ }
   \def\endpj{\relax}
   \twelvepoint\parskip=0pt\sl
   \flushright #1 \endflushenv
   \egroup\endInformal}


\def\verytop#1{
    \gdef\VERYT@P{
      \line{\hfil\vbox to 0pt{\vss\hbox{\twelvess#1}}\hfil}
      \vskip12pt }
    \gdef\verytop##1{}	}
\def\VERYT@P{\vglue-\topskip}				
\ifLocalProcessing
\else
  
\fi

\def\eprintcode#1{\gdef\@EPRINTcode{#1}}\eprintcode{}	

\def\manuscriptdate#1{\gdef\@MANUSCRIPTdate{#1}}
\manuscriptdate{\today}					
%
\def\journalname#1{\gdef\@JOURNALname{#1}}		
\journalname{}						
%
\def\manuscriptcode#1{\gdef\@MANUSCRIPTcode{#1}}
\manuscriptcode{}					


\newdimen\abstractshr@nk				
\def\MyTitleP@ge{
   \def\banner{
      \VERYT@P
      \hrule height0.2pt\vskip1.5pt\hrule height0.2pt	
      \vskip4pt\relax
      \line{\twelvepoint\rm\@LEFT \hfil \@RIGHT}
      \vskip-\baselineskip\centerline{\@CENTER}		
      \vskip3pt%
      \hrule height0.2pt\vskip1.5pt\hrule height0.2pt%
              }
   \def\title{
      \endmode\vskip 1cm plus 1fil			
      \mark{Title Page\NX\else Title Page}
      \bgroup\setparskip				
      \let\endmode=\endtitle
      \def\\{\par}					
      \center\Tbf}
   \let\@author=\author
   \def\author{\vskip0pt plus 0.1fil \@author}		
   \def\abstract{
      \endmode\bigskip\bigskip
      \vskip0pt plus .5fil				
      \centerline{A\,B\,S\,T\,R\,A\,C\,T}		
      \medskip\bgroup
      \let\endmode=\endabstract
      \narrower\narrower\normalspacing			
      \abstractshr@nk=\baselineskip			
      \advance\abstractshr@nk by -\normalbaselineskip	
      \baselineskip=\the\baselineskip minus \abstractshr@nk
      \noindent}
   \def\informalabstractskip{\vskip 0.75\baselineskip}
   \def\pacs##1{
      \vskip\baselineskip\vskip0pt plus 2fil		
      \centerline{PACS numbers: ##1}
      \smallskip}
   \def\endtitlepage{
      \vskip0pt plus 1fil				
      \preprintonly{\vskip 0pt plus 0.5fil
                    \leftline{{\tenpoint\TeXsis~\fmtversion}}}
      \endmode\eject\egroup
                    }
   \def\MyTitleP@ge{}					
                       }


\def\runningsectionskip{\bigskip}

\def\nomor@styl@s{
      \gdef\@preprint{}\gdef\preprint{}
      \gdef\@PhyRev{}\gdef\PhyRev{}
      \gdef\@paper{}\gdef\paper{}\gdef\Manuscript{}
                 }

\newif\ifpreprint\let\@preprint=\preprint
\def\preprint{
      \preprinttrue
      \def\@LEFT{\@PUBdate}
      \def\@CENTER{\@EPRINTcode}
      \def\@RIGHT{\@DOCcode}
      \@preprint\normalspacing				
      \def\Tbf{\sixteenpoint\bf}
      \MyTitleP@ge					
      \def\pacs##1{\vskip0pt plus 3fil}			
      \nomor@styl@s					
             }

\newif\ifManuscript
\def\Manuscript{
      \Manuscripttrue
      \def\@LEFT{\@MANUSCRIPTdate}
      \def\@CENTER{\@JOURNALname}
      \def\@RIGHT{Manuscript\,\#~\@MANUSCRIPTcode}
      \def\normalspacing{\lsfactor{1.75}\parskip=0pt} 
      \@preprint\normalspacing
      \def\Tbf{\sixteenpoint\bf}
      \MyTitleP@ge					
      \noInformal\FiguresLast\nomor@styl@s
               }

\newif\ifpaper\let\@paper=\paper
\def\paper{
      \papertrue\@paper\normalspacing
      \def\Tbf{\sixteenpoint\bf}
      \def\pacs##1{\relax}				
      \let\@abstract=\abstract
      \def\abstract{\@abstract\noindent}
      \def\informalabstractskip{\medskip}
      \def\endtitlepage{\endmode\goodbreak
                        \vskip .5in\egroup}
      \nomor@styl@s					
          }

\newif\ifPhysRev\let\@PhysRev=\PhysRev\def\PRcomma{}	
\def\PhysRev{
      \PhysRevtrue\@PhysRev\noInformal
      \def\scr{\fam\scrfam}			
       \let\@title=\title
      \def\title{\@title\unobeylines		
                 \def\\{\par}}			
      \def\PRcomma{,\space}			
      \def\pacs##1{
         {\leftskip=1in\noindent
          \ninepoint PACS numbers: ##1		
          \smallskip} }
      \EnvLeftskip=20pt\EnvRightskip=10pt	
      \def\quotefont{\ninerm}
      \def\quotespacing{\singlespaced}
      \SetRefs					
      \def\vol##1{{\bf ##1}}			
      \def\refFormat{\ninepoint}		
      \def\@refitem##1##2{\message{##1.}
         \refskip\noindent\hskip-\refindent
         \hbox to\refindent{\hss[##1]\SP{.3em}}
                      ##2}%
      \def\runningsectionskip{\smallskip}	
      \nomor@styl@s				
            }

\long\def\preprintonly#1{\ifpreprint#1\fi}
\long\def\Manuscriptonly#1{\ifManuscript#1\fi}

%

\long\def\Manuscriptexclude#1{\ifManuscript\relax\else#1\fi}



\def\ListReferences{
   \errorstopmode				
   \emsg{}\emsg{ \l@m Final reference list\t@m}
   \p@nctwrite.\relax
   \emsg{}
   \@refwrite{\@comment>>> EOF \jobname.ref <<<}
   \immediate\closeout\reflistout
   \begingroup\catcode`@=11\offparens\unobeylines
   \def\refskip{\vskip0pt plus 1pt}		
   \ifsuperrefs						
       \setbox\tempbox\hbox{\the\refnum.\quad}		
       \def\@refitem##1##2{\message{##1.}
            \refskip\noindent\hskip-\refindent
            \hbox to \refindent {\hss ##1.\quad}
            ##2}%
   \else						
       \setbox\tempbox\hbox{[\the\refnum]\quad}		
       \def\@refitem##1##2{\message{##1.}
            \refskip\noindent\hskip-\refindent
            \hbox to \refindent {\hss [##1]\quad}
            ##2}%
   \fi
   \refindent=\wd\tempbox \leftskip=\refindent
   \parindent=\z@ \def\reference{\@noendref}
   \refFormat						
   \Input\jobname.ref\relax\vskip0pt\endgroup\emsg{}
                   }

%
\def\checktags{
   \newlinechar=`\^^J
   \advance \@BadTags by \@BadRefs		
   \ifnum\@BadTags=\z@
   \immediate\write16{
  ^^J
  \l@m All labels and references have been successfully processed.^^J
  \l@m Please print the DVI file \jobname.dvi; afterwards, you may wish^^J
  \l@m to delete the auxiliary files \jobname.aux, \jobname.ref,
       and \jobname.log.^^J
  \l@m (In case of printing problems, see opening comments in \jobname.tex.)
  ^^J
                  }
   \else
   \immediate\write16{
  ^^J
  \l@m Attention! Don't print anything. Don't delete any file.^^J
  \l@m There were \the\@BadTags\space unresolved labels:
       you need a second run.^^J
  \l@m Please process \jobname.tex once again!
  ^^J
                  }
   \fi        }




\def\AP{Ann.\ Phys.\ (N.Y.)}		
\def\CMP{Commun.\ Math.\ Phys.}		
	\def\HPA{Helv.\ Phys.\ Acta}

\def\NP{Nucl.\ Phys.}			\def\PL{Phys.\ Lett.}

\def\PRD{Phys.\ Rev.\ D}		\def\PRL{Phys.\ Rev.\ Lett.}


\def\al{\alpha}
\def\be{\beta}
	 	
\def\de{\delta}	 	
 					  \def\vep{\varepsilon}

\def\et{\eta}
	 	  \def\vth{\vartheta}

\def\la{\lambda} 	
\def\rh{\rho}
\def\si{\sigma}	 	
\def\ta{\tau}

\def\ph{\phi}	 \def\Ph{{\mit\Phi}}		  
\def\ch{\chi}
\def\ps{\psi}	 	
\def\om{\omega}	 	


\def\sA{{\s@r A}}  \def\sB{{\s@r B}}  \def\sC{{\s@r C}}  \def\sD{{\s@r D}}
\def\sE{{\s@r E}}  \def\sF{{\s@r F}}  \def\sG{{\s@r G}}  \def\sH{{\s@r H}}
\def\sI{{\s@r I}}  \def\sJ{{\s@r J}}  \def\sK{{\s@r K}}  \def\sL{{\s@r L}}
\def\sM{{\s@r M}}  \def\sN{{\s@r N}}  \def\sO{{\s@r O}}  \def\sP{{\s@r P}}
\def\sQ{{\s@r Q}}  \def\sR{{\s@r R}}  \def\sS{{\s@r S}}  \def\sT{{\s@r T}}
\def\sU{{\s@r U}}  \def\sV{{\s@r V}}  \def\sW{{\s@r W}}  \def\sX{{\s@r X}}
\def\sY{{\s@r Y}}  \def\sZ{{\s@r Z}}


\def\etal{{\it et~al.\/\ }}
\def\circa{{\it circa\/\ }}
\def\adhoc{{\it ad~hoc\/\ }}

\def\SP#1{\hskip #1\relax}			
\def\:{\mskip -1.5mu}				

\def\D{\sD}					
\def\rd{{\rm d}}				
\def\pl{\partial}				
\def\dd#1{{{\pl\,}\over{\pl#1}}}		
\def\R{[-\infty, \infty\,]}			
\def\BZ{[-\pi\:/\:\ell, \pi\:/\:\ell\,]}	

\def\hf{{\textstyle{1\over2}}}
\def\quarter{{\textstyle{1\over4}}}

\def\intx{\int \!{\rm d}^4\:x\,}
\def\sumx{\sum_x\l^{\,4}\,}
\def\intinf{{\int\limits_{-\infty}^{+\infty}}}
\def\intint{\intinf\mskip-9.0mu\cdots\mskip-9.0mu\intinf}

\emsg{\sp@ce Loaded.}\emsg{}

\ATlock						


\texsis\preprint\lsfactor{1.15}
\pubdate{October 1993}
\eprintcode{hep-lat/9309013}
\pubcode{AKT-93-L4}


\titlepage

\title							
Weyl Neutrinos On A Lattice:
An Explicit Construction
\endtitle

\author
Anil K. Trivedi
The Enrico Fermi Institute, The University of Chicago\PRcomma
5640 South Ellis Avenue, Chicago, Illinois 60637, USA
\email{E-mail: trivedi@yukawa.uchicago.edu}
\endauthor

\abstract
Introducing a new and universally applicable discretizing technique,
I construct a class of local and unitary lattice theories of Weyl
neutrinos; this solves a longstanding and allegedly unsolvable
problem in quantum field theory. En route, I also prove a general
``go'' theorem that all Lagrangian-density based continuum quantum
field theories can be lattice-regularized.%
\informalabstract
You didn't study the Nielsen-Ninomiya theorem, only trusted the
authors to have proven the ``absence of neutrinos on a lattice''.
Well, they didn't.
Nor can anyone else: every continuum theory can be lattice-regularized.
A proof of that, plus an explicit construction of lattice neutrinos:
if you read only one paper this year, here it is!
{}From now on, this is how chiral fermions should be latticized.
All else is gaslight.\,%
\endinformalabstract
\endabstract


\endtitlepage


\def\t{\hat t}\def\l{\ell}\def\d{\rd}\def\L{\sL}
\def\psbar{\bar\ps}\def\deop{\hat\de}\def\vthop{\hat\vth}
\def\hightau{{\raise1pt\hbox{${\displaystyle\ta}$}}}\def\tauop{\hat\hightau}
\def\higheta{{\raise1pt\hbox{${\displaystyle\et}$}}}\def\etop{\hat\higheta}
\def\lowM{{\lower2pt\hbox{${\!\scriptstyle M}$}}}
\def\intpsi{\int\!\D\:\ps\,\D\:\psbar\>}
\def\intphi{\int\!\D\:\ph\,\D\:\bar\ph\>}
\def\highmu{{\raise1pt\hbox{${\displaystyle\mu}$}}}
\def\ratio#1#2{ {{\raise2pt\hbox{$#1$}} \over {\lower2pt\hbox{$#2$}}} }


The primary aim of this letter is to solve a longstanding and allegedly
unsolvable problem in quantum field theory, namely, the local and unitary
lattice-regularization of chiral Weyl~(``neutrino'') fields%
\reference{list}
While a lattice theory was first considered by
G.~Wentzel, \jnl'\HPA'13'269'1940',
their modern revival is due to
K.~G.~Wilson, \jnl'\PRD'10'2455'1974'.
Recognizing the problem with fermions,
Wilson suggested a compromise scheme in Erice lectures~(1975);
it retains locality but sacrifices chirality.
Though differently, so does the ``staggered'' formalism:
J.~Kogut and L.~Susskind, \jnl'\PRD'11'395'1975';
T.~Banks, J.~Kogut and L.~Susskind, \jnl'\PRD'13'1043'1976'.
A nonlocal scheme, similar to Wentzel's, is due to
S.~D.~Drell, M.~Weinstein and S.~Yankielowicz, \jnl'\PRD'14'1627'1976'.
\It {None of these schemes can handle local Weyl neutrinos.}
Non-existence statements include:
L.~H.~Karsten and J.~Smit, \jnl'\NP'B183'103'1981';
L.~H.~Karsten, \jnl'\PL'B104'315'1981';
J.~M.~Rabin, \jnl'\PRD'24'3218'1981'; \jnl'\NP'B201'315'1982';
F.~Wilczek, \jnl'\PRL'59'2397'1987';
A. Pelissetto, \jnl'\AP'182'177'1988'.
The best-known of the genre is the Nielsen-Ninomiya ``no-go'' theorem:
H.~B.~Nielsen and M.~Ninomiya, \jnl'\NP'B185'20'1981';
\jnl''B195(E)'54'1982';
\jnl''B193'173'1981'; D.~Friedan, \jnl'\CMP'85'481'1982'
\endreference
\reference{technical}
Technical preliminaries:
$\hbar\!=\!c\!=\!1$; the spacing $\!=\!\l$;
the metric $ \!=\! ( 1, -1, -1, -1 ) $;
$\si^\al \!\equiv\! (1,\si^i)$.
{The summation convention is automatically suspended for
indices not explicitly balanced}: $\al$~is summed
in $ A^\al  B_\al$, but not in $A_\al  B_\al $ or $B_\al^2$.
A~lattice operator $\hat\la$ is a function
of $\t_\al \!\equiv\! \t_\al^{\,+}$
and $\t_\al^{\,*} \!\equiv\! \t_\al^{\,-}$
whose right- and left-actions are
$\t_\al^{\,\pm} f(x) \!\equiv\! f( x {\pm} \l e_\al)$
and $ f(x)\,\t_\al^{\,\pm} \!\equiv\! f( x \:\mp\: \l e_\al)$.
Identity
$\sum_x(f_1\hat\la)f_2=\sum_xf_1(\hat\la f_2)$,
the lattice analog of integration by parts, shows that under sum,
operators may be taken as acting in either direction.
The conjugate~(${}^*$) of an operator is defined by exchanging
$\t_\al \!\leftrightarrow \t_\al^{\,*}$ and replacing $i\rightarrow-i$.
An operator is local if it is a polynomial in the $\t\,$'s. By ``smearing
operator'' I mean a nonconstant local operator that reduces to~$1$
in the continuum limit~$(\l \!\rightarrow\! 0)$
\endreference.\space
Towards that, I shall construct a specific class of such theories, varying
in the steepness of regularization. En route, I shall also prove that every
Lagrangian-density based continuum theory can be lattice-regularized.

The problem needs very little introduction:
The lattice is a powerful nonperturbative platform.
Neutrino-like fields, the most fundamental of all,
are indispensable to our models.
So is locality to practical work.
However, the problem of local lattice neutrinos has mainly attracted
``no-go'' theorems and is commonly regarded as unsolvable\cite{list}.

That conventional wisdom is untenable for deep mathematical and
physical reasons:
(i)~If you say that something exists in the continuum but not on the
lattice, what you are claiming in the \It{momentum} representation is
that it exists in~$\R^4$ but not in~$\BZ^4$. That is 
impossible since the latter two spaces are isomorphic, they are the
same space labeled differently.
\Ignore 
(we can reach one labeling from the other by ``Cantor-Brouwer~mappings''%
\reference{cantor}
By ``Cantor-Brouwer mapping'' I mean a one-to-one bicontinuous correspondence
between two spaces. To possess such mappings, the spaces must be of the
same cardinality (G.~Cantor) and dimension (L.~E.~J.~Brouwer)
\endreference).\space
\endIgnore 
(ii)~Neutrinos are observed in Nature.
If such particles cannot exist on lattices, it stands proven that spacetime
is not a lattice at any scale. Yet,
at any untested scale, that is an experimentally open question; a particle
observed in an apparently continuous spacetime should be \It{describable}
on lattices.

Failure to identify such theories mainly implicates our assumptions as
faulty; perhaps, as with the rethinking of paradigms needed to consistently
``cut~off'' velocities (relativity) or phase-space volumes (quantum
mechanics), the spacetime cutoff too requires thinking ``out of the box''.
But what should we change, where, and how?
A methodological contribution of this paper is to identify one viewpoint
in which an exact solution is easily apparent.

I shall translate the existence of isomorphisms between the two momenta%
\reference{topology}
The usual convention of viewing the two spaces through different
topologies is not relevant here
\endreference\space
into a discretizing technique. It is not an \adhoc approach to the lattice
neutrino problem, but will apply to a variety of structures%
\reference{tobe}
A.~K.~Trivedi, to be published
\endreference,\space
including \It{all} theories that develop species-doubling upon usual
discretization%
\reference{doubling}
These range from the time-dependent Schr\"odinger equation to
solitons~[J.~Govaerts, J.~Mandula, and J.~Weyers, \jnl'\PL'112B'465'1982']
to gravity~[P.~Menotti and A.~Pelissetto, \jnl'\PRD'35'173'1981']
\endreference.\space
However, I shall limit this discussion to quantum field theories,
and the construction itself to ultraviolet-regularized lattice
theories of neutrinos (unregularized lattice theories can exist%
\reference{l1}
A.~K.~Trivedi, \jnl'\PRL'61'907'1988'
\endreference;\space
please note the perimeter of regularization when I identify it).
I shall not discuss the doubling\cite{tobe}: the goal here is not
so much to understand that unwanted phenomenon as to simply avoid it.
This work confirms that problems with lattice fermions are due to
{overlocalization} which needs ``smearing out''%
\reference{l2}
A.~K.~Trivedi, \jnl'\PL'B230'113'1989'
\endreference:\space
to paraphrase Einstein, physics should be made as local as possible but
not any more local. This work is independent of the Nielsen-Ninomiya
theorem\cite{list} whose critique has been given separately%
\reference{l3}%
\Manuscriptexclude{A.~K.~Trivedi, hep-lat/9309012}%
\Manuscriptonly{Physical Review Letters manuscript LJ5263 \It{(the Editor
is requested kindly to add the reference)\rm}}
\endreference.\space

Here is the arena. A physical {theory} predicts certain observables.
The choice of a working parameterization is called a {representation}.
A lattice quantum field theory consists of instructions like
$$
   Q^A(u') \equiv N\int [\D\ch]_\rh
             \> q_\rh^A \! [ \ch\! ( u' ) ]
             \> \exp i S _\rh [ \ch\! (u) ] ,
\EQN QA $$
where $\ch$ denotes all fields
and $N \:\equiv\: 1\,/ \:\int[\D\ch]_\rh \,\exp iS_\rh [\ch]$.
Both $\ch$ and $u$ are dummy variables:
$\ch$~parameterizes the path integral and defines the ``outer''
or ``functional'' representation $\rh$ of the theory; $u$ parameterizes
the domain of $\ch$, over which the ordinary integral (or sum)
$\,S_\rh [\ch] \!=\! \int \d u \,\L_\rh[\ch(u)]\,$ is evaluated,
and defines the ``inner'' representation. Being dummy, both can be changed.
Changing path variables yields representations differing in both the
action $S[\ch]$ and the path-transcription $q^A(\ch)$ of
observables\cite{l3}---notice that one has no physical relevance without
the other. A representation is
(i)~local if it preserves the locality of the continuum action and
observables; I shall call it
(ii)~``hamiltonian'' if the Lagrangian-density has a continuous-time
limit of the form $\ch[i\dd{t} \!-\! \hat H]\ch $, and (iii)~``canonical''
if the expressions $q^A(\ch)$ coincide with their continuum counterparts.
In the compact formulation of the theory,
$$
  W[J] = N \int[\D\ch]_\rh \;\exp iS_\rh [\ch;J],
\EQN WJ $$
where the integrals \Ep{QA} arise as the coefficients in the
functional Taylor-Berezin series in $ J $,
a representation $\rh$ is (i)~local if $S_\rh [\ch;J]$ is local,
(ii)~hamiltonian if $S_\rh [\ch;0] $ obeys the above mentioned
continuous-time limit, and (iii)~canonical if the source-field coupling
is $ J\ch $.
(Thus a theory is determined not by a spectrum,
an equation, or an action,
but by its observables, the functional derivatives of $W[J]$ at $J=0$.
Just as in transitions to relativity and quantum
mechanics, it is necessary to alter the mathematical transcription of
the observables along with the dynamics: the essential result here will
be that regularized local unitary lattice theories of neutrinos exist,
but the representations which manifest this locality are neither
hamiltonian nor canonical.)

After a decade of ``no-go'' theorems, you may find it refreshing
to encounter a ``go''~theorem:
\It{Every Lagrangian-density based continuum quantum field theory
can be lattice-regular\-ized}.
(That covers {all} fundamental theories presently envisioned.)
Here is a terse but complete proof.
Let $J$ denote all sources, $q_\al \!\in\! \R$ the coordinates,
and $p_\al \!\in\! \R$ the momenta.
A theory would be defined by a functional $W_0 [J(q)]$ in the coordinate
and $W_0 [J(p)]$ in the momentum representation. Take the latter.
By assumption, it can be written
as $W_0 \!=\! N\!\int\!\D\ch \exp i\!\int\!\highmu_0\,\L_0$
where $\highmu_0 \!\equiv\! \d^4p/(2\pi)^4 $.
It can be modified to obtain others,
$W_\l \!=\! N\!\int\!\sD\ch \exp  i\!\int\!\highmu_\l\,\L_\l$,
where $\l$ is a parameter, $\highmu_\l\:=\:\xi(\l;p)\,\highmu_0$,
and $W_\l \!\rightarrow\! W_0$ as $\l \!\rightarrow\! 0$; in particular,
a regulator $\xi(\l;p)$ can {always} be found to suppress the measure
in the ultraviolet (or any other) region to any specified degree.
{}From the regularized functionals $W_\l$, 
take a specific $W[J(p)]$.
Any one of the infinitely many isomorphisms $\la_\l$ between the real
line and Brillouin Zone yields a $W[J(k)]$ where $k_\al \!\in\! \BZ$.
By Fourier analyzing to the coordinate representation conjugate to~$k$,
we obtain $W[J(x)]$, i.e., a lattice\ theory%
\reference{proof}
While this is undoubtedly a ``physicists's proof'', its mathematical
ingredients---Fourier analysis; existence of functions $\xi(\l;p)$ which
decrease arbitrarily fast for large $p$; mappings between real intervals
and the real line---are all well established, and it can be made as
rigorous as you desire
\endreference.\space

This proof shows that each continuum theory possesses lattice
\It{representations} (equivalent to it)%
\reference{CL}
This may be referred to as ``Continuum-Lattice Duality''
\endreference,\space
as well as \It{transcriptions} (equivalent in the $\l \!\rightarrow\! 0$
limit); typically one has a divergent continuum theory and seeks a
regularized lattice transcription.

I note a few practical aspects of this admittedly spartan existence argument:
(i)~Even if $W[J(p)]$ can be studied (``is\ defined'') only perturbatively,
$W[J(x)]$ can be subjected to nonperturbative lattice tools.
(ii)~Besides regulating the measure ($\highmu_0\!\rightarrow\!\highmu_\l$),
you can use the modification $\L_0 \!\rightarrow\! \L_\l$ to fine-tune
other properties if desired.
%
%
(iii)~The theorem can be made stronger:
\It{unitarity- and locality-preserving transcriptions always
exist}\cite{tobe}.
I shall present the general argument elsewhere; for neutrinos,
unitary local theories are obtained below by explicit construction.
(iv)~The theorem suggests a practical discretizing technique: use an
isomorphism $\la_{\l}$
to associate a continuum-like space with the lattice; write theories in it;
choose convenient field variables; change the inner variables back to~$x$.

One such construction of lattice neutrinos is presented below in
four steps.

First, I associate a continuum with the lattice:
\def\gap{\SP{1.25em}}	
\def\denoma{{1\:+\:\quarter\l^2p_\al^2}}
\def\denomb{{1\:+\:\quarter\l^2p_\be^2}}
$$
p_\al =  {2\over\l} \tan \hf\l k_\al,  	\gap
p_\al \!\in\! \R, 		    	\gap
k_\al \!\in\! \BZ .
\EQN p.def $$
It is convenient to introduce the functions:
$$
\et\:(p) \:=\: \prod_{\al=0}^3 {{1\:-\:\hf i\l p_\al}\over\denoma} ,	\gap
\vth\:(p) \:=\: \prod_{\al=0}^3 {1\over\denoma} ,			\gap
\de_\al\:(p) \:=\: {p_\al\over\denoma}\prod_{\be\neq\al}{1\over\denomb} ;
\EQN defs;a  $$
their conventional representations:
$$
\et\:(k) \:=\: \prod_{\al=0}^3  \hf (1 \:+\: e^{-i\l k_\al} \: ), \gap
\vth\:(k) \:=\: \prod_{\al=0}^3  \cos^2 {\hf}\l k_\al   ,         \gap
\de_\al\:(k) \:=\: {1\over\l}
             \sin\l k_\al \prod_{\be\neq\al} \cos^2\hf\l k_\al ;
\EQN defs;b $$
and their operator representations:
$$
\etop    \:=\: \prod_{\al=0}^3 \hf (1 \:+\: \t_\al) , 		          \gap
\vthop    \:=\: \prod_{\al=0}^3 \quarter(\t_\al\:+\:2\:+\:\t_\al^{\,*}) , \gap
\deop_\al \:=\:  {i\over{2\l}}  (\t_\al\:-\:\t_\al^{\,*})
	     \prod_{\be\neq\al} \quarter(\t_\be\:+\:2\:+\:\t_\be^{\,*}).
\EQN defs;c $$
%

Second, I choose a Lagrangian-density in $\{p\}$-space:
%
$$
 \L_\l= \L_0 = \bar\ph  \,\si^\al  p_\al\,  \ph  +
         \bar J \ph  +  \bar\ph J .
\EQN L $$
It determines a classical theory identical with the continuum Weyl theory.
There is \It{no~question} of doubling and, using \Ep{p.def}, the spectrum
$p^\al  p_\al \!=\! 0$ can be written as
$$
 \tan^2\hf\l\om-\sum_{i=1}^3 \tan^2\hf\l  k^i =0 ;
\EQN spectrum $$
all quantum theories to follow possess this spectrum.

Third, I quantize {and regularize} the theory. That requires an action
$S\!=\!\int\highmu_\l\,\L_\l$ where the measure $\highmu_\l$ vanishes suitably
fast as $p_\al\!\rightarrow\!\pm\infty$. While the standard lattice measure
$\d^4k \!=\! \vth\,\d^4p$ is certainly adequate, consider more generally:
$$
  \highmu_{\l,M} = \vth^M \, \d^4\:p /\!(2\pi)^4 , \quad M = 1,2,3,\ldots
\EQN measure $$
where $M\ge1$ is a positive-definite integer ($M\!=\!1$ gives
the standard case). This yields
$$
 W_{\l,M} [\bar J,J] =  N\!\intphi\exp i\!\intint
\left\{ \prod_{\al=0}^3 {{\d p_\al/2\pi} \over {(\denoma)^\lowM}} \right\}
\! (\bar\ph\si^\be p_\be\ph + \bar J \ph +\bar \ph J) .
\EQN Wphi $$
The regularization here improves with increasing $M$ since
$\highmu_{\l,M} \propto p_\al^{-2M}$ as $p_\al\!\rightarrow\!\pm\infty$;
along a generic direction in the euclideanized space,
the measure $\highmu_{\l,M}$ falls as $|p|^{-8M}$.
(In $n$~dimensions it falls as $|p|^{-2nM}$; overcoming a classic continuum
limitation, the theory now remains regularized in higher dimensions.)

Before proceeding further, you should satisfy yourself---the familiar
continuum tools suffice---that for $M \!\ge\! 1$, \Eq{Wphi} does define
ultraviolet-regularized quantum field theories of neutrinos;
from now on, it is only change of variables.

Fourth and the last, I bring forth the hidden lattice parameterization.
The following yields the simplest local and manifestly hermitian
representations: First, writing $M\!=\!2m+\vep$,
where $\vep\!=\!M\mod2$ (i.e., $0$ if $M$ is even; $1$ if $M$ is odd)
and $m\!=\! {\rm Int}\,(M/2)$,
change path-variables (outer representation) to
$$
\ps= 	\vth^{\,m-1}	\,	\et^{*\vep}	\,	\ph,	\qquad
\psbar=	\vth^{\,m-1}	\,	\et^\vep	\,	\bar\ph;
\EQN transform $$
the Jacobian $\D\ph\D\bar\ph\,/\,\D\ps\/\D\psbar$ is a functional constant.
Next, change the inner representation to $x$; this means using \Eq{p.def}
to revert the momentum space parameterization to~$k$ (the Jacobian is
$\d^4p/\d^4k \!=\! \vth^{-1}$) and going to the coordinate representation.
The result is
$$
W_{\l,M} [\bar J,J] =
N \intpsi \exp i\sumx [ \psbar\si^\al\deop_\al\ps  +
\bar J \, \tauop_\lowM\ps+\overline{\tauop_\lowM\ps} \, J ],
\EQN Wpsi $$
where $\tauop_\lowM \!=\! \vthop^{\,m\,} \etop^\vep$
(said differently, $\tauop_{2m} \!=\! \vthop^{\,m\,}$
and $\tauop_{2m+1} \!=\! \vthop^{\,m} \etop$; the first few
$\tauop$'s are $\tauop_1 \!=\! \etop$, $\, \tauop_2 \!=\! \vthop$,
$\, \tauop_3 \!=\! \vthop\etop$, and so on).

We now have a class of lattice theories of neutrinos:
Each positive-definite integer $M\!\ge\!1$ defines a distinct theory.
In the chosen representation, all $M$-dependence is in the
source-field coupling, the action is $M$-independent.
The ultraviolet regularization improves with increasing $M$.

The formalism \Ep{Wpsi} lattice-transcribes a continuum $q^A(\psbar,\ps)$
as $q^A(\overline{\tauop_\lowM\ps},\,\tauop_\lowM\ps)\,$%
\reference{example}
For example, the continuum $\psbar\si^\al\ps$
will be lattice-transcribed as
$(\overline{\tauop_\lowM\ps})\si^\al(\tauop_\lowM\ps)$.
Indeed, formalisms \Ep{Wpsi} admit no observable like~$\psbar\si^\al\ps$.
This should not disturb you: just as quantum mechanics can alter
mathematical representation of classical observables, a lattice
theory too need only transcribe, not retain, continuum expressions
\endreference\,
and then averages it with the action $\sumx\psbar\si^\al\deop_\al\ps$:
\It{ i.e., each continuum path-average
$$
Q_0^A = \ratio%
{\intpsi\;q^A\lparen\psbar,\ps\rparen\;\exp i\intx\psbar\si^\al i\del_\al\ps}%
{\intpsi \exp i\intx\psbar\si^\al i\del_\al\ps}
\EQN rule;a $$
is lattice-regularized as}
$$
Q_{\l,M}^A = \ratio%
{\intpsi\;q^A\lparen\,\overline{\tauop_\lowM\ps},\tauop_\lowM\ps\rparen\;
	\exp i\sumx\psbar\si^\al\deop_\al\ps}%
{\intpsi\exp i\sumx\psbar\si^\al\deop_\al\ps} .
\EQN rule;b $$

Equation \Ep{Wpsi}, or its restatement \Ep{rule}, is the main result
of this paper. It defines a class of manifestly local, unitary,
translationally invariant lattice theories;
that they are regularized theories of neutrinos is manifest in the
representation \Ep{Wphi}.

These formalisms possess two unusual features:
(i)~Derivatives $\deop_\al$ incorporate transverse
smearing; this gives the action a non-hamiltonian structure.
(ii)~The ultraviolet regularization [$M \!\geq\! 1$, \Eq{Wphi}] further
requires non-canonical source-field coupling [\Eq{Wpsi}], i.e., smeared
transcription of observables [\Eqs{rule}]. In other words, the formalisms
do not incorporate the continuum Born interpretation but a smeared
version of it. This smearing involves quantum interference of nearby
fields---notice that \Eq{Wpsi}, or \Ep{rule;b}, automatically generates
the Schwinger ``point-splitting'' which upon gauging leads to chiral
anomaly---rather than classical averaging of observables.
Both of these features are simultaneously needed%
\reference{smearing}
This derives the 	
suggestion of \Ref{l2} that the ultraviolet problems plaguing lattice
fermions can be solved by smearing out two instances of overlocalization,
each incompatible with the limited spacetime resolution a fundamental
length mandates:
(i)~the infinitely sharp transverse localization of differentiation, and
(ii)~the possibility of observations limited to a single point
\endreference.

Incidently, the only relevance that the attempted ``no-go'' arguments
offered in the Nielsen-Ninomiya theorem\cite{list} have for this subject
is as follows: If in \Eq{Wpsi} or \Ep{rule;b} you change path variables
so as to make the action hamiltonian, it will also become nonlocal in that
parameterization\cite{l3}. Since parameterizations do not affect observables
and are chosen for convenience, this has no importance beyond telling us
that such parameterizations should be avoided in practical computations.

Equations \Ep{Wphi} and \Ep{Wpsi} yield a useful rule: \It{with the action
held fixed, greater smearing of~$\ps$ in the source-field coupling, i.e.,
in the lattice transcription of the observables, improves the ultraviolet
regularization}.
The same pattern holds in the interaction with a dynamical field:
For example, the simplest ultraviolet-admissible Yukawa interaction
(with a scalar $\Ph$ possessing the standard kinetic term) is
$\Ph\;\overline{\etop\ps}\;\etop\ps$; the interaction
$\Ph\;\overline{\vthop\ps}\;\vthop\ps$  leads to better
regularization; and so on.

The simplest one of these formalisms is
$$ \EQNalign{
W_{\l,\,1} [\bar J,J]
&= N \intpsi \exp i\sumx [ \psbar\si^\al\deop_\al\ps  +
\bar J \etop\ps+\overline{\etop\ps} J ],
\EQN W1.herm
\cr
Q_{\l,\,1}^A
&=  \ratio%
{\intpsi q^A\lparen\,\overline{\etop\ps},\etop\ps\rparen\,
	\exp i\sumx\psbar\si^\al\deop_\al\ps}
{\intpsi\exp i\sumx\psbar\si^\al\deop_\al\ps} ;
\EQN rule.herm
\cr} $$
$\etop$ and $\deop_\al$ are defined in \Eq{defs;c}. Although I would
recommend the manifestly hermitian representation \Ep{W1.herm}--\Ep{rule.herm},
the representation \It{can} be simplified further by giving up the
\It{manifestness} of hermiticity (I omit the details):
\def\highplus{{\raise1pt\hbox{${\scriptstyle+}$}}}
$$ \EQNalign{
W_{\l,\,1} [\bar J,J]
&= N\intpsi \exp i \sumx
    [\psbar\si^\al\hat\de^\highplus_\al\ps + \bar J \etop\ps + \psbar J] ,
\EQN W1.nonherm
\cr
Q_{\l,\,1}^A &= \ratio%
{\intpsi q^A\lparen\psbar,\etop\ps\rparen\,
	\exp i\sumx\psbar\si^\al\hat\de^\highplus_\al\ps}
{\intpsi\exp i\sumx\psbar\si^\al\hat\de^\highplus_\al\ps} ,
\EQN rule.nonherm
\cr
\hat\de^\highplus_\al  &= {i\over\l} (\t_\al \:-\: 1)
	\prod_{\be\neq\al} \,\hf (\t_\be \:+\: 1) ;
\EQN deltaop
\cr} $$
if you opt for this simplification, expressly ensure unitarity when
adding interactions.

Explicitly, the Lagrangian-density in \Eq{W1.herm} has $248$ terms contributed
by $81$ sites constituting the $3^4$ hypercube centered at~$x$; that in
\Ep{W1.nonherm} has $81$ terms from $16$ sites constituting the $2^4$
hypercube with a corner at~$x$.
You could complain that these formalisms will strain today's computers,
but one could just as reasonably have faulted Maxwell's theory or general
relativity for being too complex for slide rules.
Physical problems often have an irreducible complexity we must face;
I shall show elsewhere that there is no simpler local unitary latticization
of  neutrinos than $W_{\l,\,1}$\cite{tobe}.

While the focus here was on the lattice regularization, you can cast
these theories in local $\{q\}$-space representations as well ($q$ being
the coordinate conjugate to $p$); that gives a continuum regularization
of neutrinos\cite{tobe}. [However, the measure \Ep{measure} was specifically
optimized for the lattice. If it is the continuum regularization that
you want, other choices will be more appropriate.]

I promised to identify the perimeter of the ultraviolet-regularization.
{}From \Eq{Wphi}, we need $M\!\ge\!1$. Basically, all that is asked of
us is not to sleepwalk into the $M\!=\!0$ case:
\It{if you merely latticize the Lagrangian-density as
$\psbar\si^\al\deop_\al\ps$ but retain the continuum transcription
$q(\psbar,\ps)$ of observables, you get a lattice theory which,
though free from the doubling, is not regularized}. You would not
expect it to yield the anomaly, for example. I cannot possibly
overemphasize the following: \It{if you want locality, unitarity,
chirality, and the regularization all at once, then don't just
latticize the action but follow the prescription~\Ep{rule} completely}.
[By the same token, in judging the proposal you should examine its
final observables \Ep{rule;b}, not some part of some path integrand.]

As for the latticizing technique, I began with a continuum-like
parameterization of the lattice momentum space because that places us
in the only manifold where we know how to avoid the doubling. It lets
us explicitly control the ultraviolet regularization, and separates
the regularization from the latticization%
\reference{slac}
You may wish to compare this approach with that of Drell \etal (\Ref{list})
who also start out in the momentum space.
Those authors set the measure $\mu\!=\!0$ outside a box (Brillouin Zone),
which regularizes and latticizes at the same time, but also destroys
locality. In my approach, ultraviolet contributions are first regularized
smoothly with $\mu \!\propto\! \vth^M$  [\Eq{measure}]
and the resulting theory, already regularized, is latticized not by cutting
off the momentum space, but by mapping it to Brillouin Zone [\Eq{p.def}]
\endreference.\space
Locality could be preserved here because the regulator is a rational
algebraic function of $p_\al$ [\Eq{Wphi}] which in turn is a rational
algebraic function of~$\exp\pm{}i\l{}k_\al$ [\Eq{p.def}];
\It {this is a general theorem}\cite{tobe}.
Similar construction can discretize any continuum theory; I shall state
the equivalent final prescription elsewhere\cite{tobe}.

The construction and the go-theorem given here are but elementary applications
of a deeper principle which may be called ``Continuum-Lattice Duality'':
\It{those two mathematically distinct spaces are interchangeable as domains
of physical theories}\cite{tobe}.
This duality is analogous to wave-particle duality (which too asserts that
two mathematically distinct concepts are interchangeable as vehicles for
physical laws); it too has similarly deep consequences for physics which I
hope to discuss elsewhere.

I conclude by recalling what I did here. I showed that all
Lagrangian-density based continuum theories can be lattice-regularized;
I stated that unitarity and locality can be preserved in the process.
I introduced a new latticizing technique; I promised to spell out the
equivalent final algorithm elsewhere.
For neutrinos, I constructed a class of manifestly local, unitary,
and ultraviolet-regularized lattice theories. They do come with two
unfamiliar features (the action is non-hamiltonian, and observables
are transcribed differently from the continuum), but you can hardly
insist that introducing a new fundamental constant in physics should
not change anything.


\nobreak\bigskip\nobreak
\line {\hfil \hbox to 2.25in {\hrulefill} \hfil}
\bigbreak\ListReferences


\pj 
I do not know what I may appear to the world;\\
but to myself I seem to have been only like a boy who kept getting showered\\
with smooth and rough pebbles on the seashore,\\
whilst the great ocean of truth lay undiscovered before us all.\\
\pjauthor ANONYMOUS \rm(\circa 1675)
\endpj


\bye